# Immobility of isolated swarmer cells due to local liquid depletion

Ajesh Jose[1], Benjamín Pérez-Estay[2], Shira Omer Bendori[3], Avigdor Eldar[3], Daniel B. Kearns[4], Gil Ariel[5], and Avraham Be'er[6,7]

[1]*The Albert Katz International School for Desert Studies, The Jacob Blaustein Institutes for Desert Research, Ben-Gurion University of the Negev, Sede Boqer Campus 84990, Midreshet Ben-Gurion, Israel*

[2]*Laboratoire PMMH-ESPCI Paris, PSL Research University, Sorbonne University, University Paris-Diderot, 7, Quai Saint-Bernard, Paris, 75005, France.*

[3]*The Shmunis School of Biomedicine and Cancer Research, Faculty of Life Sciences, Tel Aviv University, Tel Aviv, 69978 Israel.*

[4]*Department of Biology, Indiana University, Bloomington, Indiana 47405, USA*

[5]*Department of Mathematics, Bar-Ilan University, Ramat Gan 52900, Israel*

[6]*Zuckerberg Institute for Water Research, The Jacob Blaustein Institutes for Desert Research, Ben-Gurion University of the Negev, Sede Boqer Campus 84990, Midreshet Ben-Gurion, Israel*

[7]*Department of Physics, Ben-Gurion University of the Negev 84105, Beer-Sheva, Israel*

## Abstract

Bacterial swarming is a complex phenomenon in which thousands of self-propelled rod-shaped cells move coherently on surfaces, providing an excellent example of active matter. However, bacterial swarming is different from most studied examples of active systems because single isolated cells do not move, while clusters do. The biophysical aspects underlying this behavior are unclear. In this work we explore the case of low local cell densities, where single cells become temporarily immobile. We show that immobility is related to local depletion of liquid. In addition, it is also associated with the state of the flagella. Specifically, the flagellar bundles at (temporarily) liquid depleted regions are completely spread-out. Our results suggest that dry models of self-propelled agents, which only consider steric alignments and neglect hydrodynamic effects, are oversimplified and are not sufficient to describe swarming bacteria.

Introduction

Bacterial swarming is a collective mode of motion in which rod-shaped cells, powered by multiple flagella, move and migrate on surfaces by forming dynamic clusters that continually split and merge[1-12]. The clusters have the structure of whirls and jets that are visually reminiscent of collective motion patterns such as seen in flocks of birds and schools of fish[13]. However, for bacteria, swarming is more than a collective movement pattern or dynamical phase, but a biological state, termed "lifestyle". Bacterial swarming has been described as a particular phenotype, in the sense that swarmers show a significant increase in the expression of some swarming-related genes[11–16], even in species that are described as swarmers but are powered by motive organelles that are not flagella. Swarming as a phenomenon is not species specific, although the cell shape in the colony, typical speeds, densities and the thickness of colonies depend on the species and the environment[1-9]. The evolutionary advantage of swarming is not completely understood, but swarmers are more resistant to antibiotic stress[19–22] and to a variety of adverse environmental conditions[23,24]. In addition, it has been hypothesized that foraging by swarmers is more efficient compared to swimmers[26].

Physically, bacteria live at low Reynolds numbers where viscous forces dominate[5,9,35,27–34]. In order to overcome viscosity and the friction between the cell and the surface, swarming bacteria secrete osmotic agents to extract water from the surface[36,37], as well as surfactants or lubricants[2,23,38,39]. Thus, it is thought that the cells inhabit a thin layer of liquid, of the order of a few micrometers, in which they eventually move. The collective cellular motion obtained during bacterial swarming has been attributed to steric alignment (mainly due to excluded volume), and to hydrodynamic interactions. Swarming is different from swimming (in bulk liquid) in many physical aspects including cell aspect ratio[40,41], the mechanism for propulsion (using a larger number of flagella when swarming)[42,43] and the quasi two-dimensional topology of swarm colonies[44].

The dynamics of both swimming and swarming bacteria have been analyzed extensively using the tools of statistical physics[9,27,29,33–35,45,46]. For example, spatial and temporal correlation functions have been used, both in simulations and experiments, to infer putative phase diagrams, describing qualitatively distinct dynamical regimes[30,31,47]. One of the prominent differences between swimmers and swarmers is their behavior at low densities. Isolated swimming cells move at relatively high speeds, of the order of 10 μm/s[33–35]. In this respect, the behavior of large populations of bacterial swimmers is similar to other examples of active systems, such as larger animals (e.g. birds, fish[13]), synthetic material (e.g. Janus particles, Quincke rollers or hexbugs[48–50]) and most simulations of self-propelled agents[51–53]. In contrast, in general, isolated swarming cells do not move move[26] and speed increase with density (up to very high densities[5,9,31,55,56]. At low to intermediate densities, this leads to an interesting phase in which large, dense moving clusters are surrounded by scattered stationary cells[5,9,31,56].

The physical mechanisms underlying the inability of isolated cells to move are currently unknown and are the main focus of this manuscript. Previous work[57] speculated that the absence of motion at low densities may depend on the lack of physical contact between cells, or the local absence of moist or water, suggesting that solitary individuals are unable to overcome the frictional forces between the cell and the surface.

Experimental setup

We work with *Bacillus subtilis* (wild-type 3610; labeled red, strain AE4847 pAE1222-LacA-Pveg-R0_mKate#2 mls, amp), which is a model swarm species that has been used extensively in the past[2,7]. The cells, which are rod-shaped ($1\times7$ μm), are grown in Petri-dishes under standard swarm conditions (0.5% agar and 25 g/l LB; incubated at 30°C, 95% relative humidity (RH)). Prior to inoculation, the plates are aged for 24 h in the lab (24°C and 30% RH) with the lid on, and then are opened for 8 more minutes to remove extra moisture. Moisture is removed to allow proper absorption of the 5-μl drop of log-phase liquid culture by the agar. Starting from a small spot at the center of the colony, with $\sim 5\cdot 10^5$ cells, the colony grows and expands rapidly outwards; we focus on the outer regions of the colony near its edge where the averaged surface fraction $\rho$ covered by cells is $\rho=0.3$. Optical microscopy was used to track the location of the cells as a function of time, from which we generate all the data (Zeiss Axio Imager Z2 at 63× with phase-contrast, brightfield and fluorescence (filter set 20 Rhodamin shift free: Excitation 546/12; Beam Splitter 560; Emission 607/80) modes, hooked to a Neo (Andor) camera operated at 1000×1000 pixels and up to 50 frames/s.

In addition, we use differential interference contrast (DIC) microscopy, which enables, among other things, the detection of surface roughness by showing different colors for different slopes in the sample (see appendix 1 for more details). We use a Zeiss Axioscope 5 at 63× to which we have added an additional Rochon prism in some of the experiments. The microscope is hooked to a color Axiocam 208 camera operated at 1000×1000 pixels with still images taken at 0.02 s exposure and video with a frame rate of 30 frames/s.

See appendix 2 for details on staining the flagella. See appendix 3 for details on image analysis.

Results

Figures 1a-b show an example of the region of interest, with the cells seen on the agar, for two different exposure times (phase-contrast). At high frame rate (50 frames/s; exposure time=0.02 s), all the cells ($\rho$=0.3) are clearly seen (Fig. 1a), but at a slower rate (5 frames/s; exposure time=0.2 s) only the cells that are temporarily immobile are seen (Fig. 1b), while the moving ones are smeared. By thresholding the image, the fraction of immobile cells is determined. To understand why some cells are *temporarily* immobile, we first checked whether the agar is smooth and uniform, as perhaps there are regions that tend to trap, or un-trap cells. The field of view is thus divided into 100 equally sized bins (10×10), and the fraction of time each bin is inhabited with an immobile cell is measured for all bins (at least 1/3 the size of a cell in a bin is needed to count one). We repeat this measurement for a variety of durations, from 5 s to 360 s in steps of 5 s. Our results show (Fig. 2a) that on average, each bin is occupied with immobile cells about 6% of the time. This result does not depend on the duration. We also look at the distribution among the 100 bins and obtain that the occupation counts are Gaussian. The standard deviation (Fig. 2a inset) decreases as $t^{-1/2}$ indicating that the stagnation of cells follows the central limit theorem. Additionally, we calculate the time correlation function, which quantifies how long the fluctuations with respect to the average occupation of 6% are maintained over time. We observe that the correlation drops to zero after just a few seconds (Fig. 2b). These results suggest that there are no fixed locations that tend to trap or un-trap the cells (Figs. 2c-d). In other words, over time, the surface is homogeneous.

In Figs. 2e-h we show results for the same experiment, only that now we have dried the agar plates for 4 times longer prior to inoculation, making the surface less wet. The results show that the number of stationary cells at a given time is larger, as the average inhabitation of bins by the immobile cells is 13% of the time. As in wetter conditions, the distribution remains Gaussian and the time correlation drops to zero, suggesting that the lack of liquid does not create fixed local traps. Overall, the results suggest that there are wet points of tiny "ponds" but those are not fixed in time and space.

Figure 3a shows results for the distribution of speeds in the two (wet and dry) cases (data was taken at 50 frames/s). We suggest that the drying time affects mostly the stationary, and the slowly moving cells (e.g., cells that move slower than 7 µm/s). Clearly, the number of non-moving cells is larger at the dry case; however, the cells that are moving faster than 7 µm/s (i.e., cells that were not affected by the lack of water) exhibit the same average speed (19.5 µm/s for the wet case and 18.6 µm/s for the dry case), and a similar speed distribution (Fig.

3b). As in past studies, the speed distribution has a Gaussian tail as expected for the Rayleigh distribution (e.g.,[22]).

In Fig. 4 we compare results obtained under different illumination techniques. At a large enough frame rate (50 frames/s), when the cells are not smeared due to their motion, the fluorescence mode, the bright-field and the phase-contrast, all show that both the stationary cells and the moving ones exhibit a similar intensity, where it is impossible to tell by a single snapshot which cells are stationary, and which ones are moving. However, the DIC image shows that stationary cells and moving ones have a different color (Fig. 4d). In DIC the color (e.g., the hue of the color obtained in the HSB color system), indicates a slope in the surface of the sample with respect to the parallel microscope table (i.e., perpendicular to the beam)[58]. The same hue is obtained for regions in the agar that are free of cells, and for the moving cells. This suggests that the local surface of the moving cells is parallel to the agar, presumably due to the presence of a thin liquid layer on top of them, causing the cells not to be easily resolved. However, stationary cells are nicely resolved and their hue reflects the slopes of their rod-shaped structure.

To increase the DIC hue-contrast even more (Fig. 5a), we have added an *additional* Rochon prism to the DIC system, and then threshold the image by the hue (we "label" all pixels with hue equal or smaller than 20 in black, in between 21-24 blue, and 25-27 in green, while the agar background, as well as the moving cells are typically in the range of 43-45). Also, there are very few pixels in the entire field of view in the range of 28-42 so that the moving cells and the stationary ones are clearly distinguished (Fig. 5b). Figures 5c-g show the gradual stopping process of a single cell. Figure 5c shows a moving cell, that then stops (Fig. 5d), and starts losing the liquid around it (Fig. 5e-g), with more regions of the cell appear to be blue and black, indicating the loss of the liquid that was covering it. Figures 5h-l show the gradual acceleration of a single cell (from stationary to moving). Figures 5h-j show a stationary cell that gets wetter as the blue and black colors disappear. In the next frame (Fig. 5k), the cell is not moving yet. It eventually accelerates (Fig. 5l) as enough of the liquid covers it. Additional results (Movie S1 and Movie S2) show tiny (~0.5 μm in diameter) MgO beads deposited on the colony (See methods in[38,59]). While beads on the virgin agar are immobile (Movie S1), those near stationary cells (Movie S2) do move, indicating the presence of liquid streams that are sufficient to drag the beads but not the cells. This establishes that liquid flow around a cell is not sufficient to move it.

Next, we study the flagellar arrangement using fluorescence staining of the live bacteria (see appendix 2 for details). The results show that swarming cells may be largely divided into three categories (Movie S3). This classification is independent of the drying time. (i) Closed (Fig. 6a); with all flagella tightly wrapped around one of the poles, (ii) (Fig. 6b) Partially open; with flagella partially unbundled, and (iii) (Fig, 6c) Open; where the flagella are completely unbundled, but are still rotating. Analyzing the trajectories of the cells, the state of the flagella is labelled manually. We obtain the probability distribution of two characteristics for each of the categories. In Fig. 6d we show the probability of speeds for the wet case (the drier case is *qualitatively* similar). Cells with completely unbundled (open) flagella (data in red) are almost never moving, while cells of the two other states do move, but with different speed distributions (inset of Fig. 6d). We also show the time durations the cells spend in each state (Fig. 6e). The data of the stationary cells is compared to the data taken from the DIC experiments with a fair similarity. We refer to the closed flagella state as "run", to the partially open (partially unbundled) state "tumble", because cells change their direction of motion and do not move in straight lines, and to the open flagella "stall".

Lastly, we establish that the transition between the three flagellar states is statistically consistent with a continuous time Markov chain. Fig. 6f shows the histogram of the waiting times in the closed and partially open states on a semi-log plot along with the maximum likelihood fit to an exponential distribution. The fit, with averages 0.25 s and 0.21 s, respectively, is very good. Fitting the duration of the closed state (Fig. 6g) to an exponential distribution (average 0.95 s), is not as good. Fitting to a mixture of two exponential distribution is better, but is not statistically justified AIC of a single exponential – AIC of mixture = -2.5). Thus, within the data available, the approximation to an exponential waiting time is justified. Figure 6h shows the transition matrix between the states. No events of shifting directly from closed (run) to open (stall) or vice versa were observed (>1000 measurements). The transition from the partial (tumble) state to run or tumble are close to equal. In order to see if these transitions are memory less, we compare the two-step transition matrix (Fig. 6i) with the square of the single step one (Fig. 6j). The two are equal up to a relative error of ~1%. Overall, we conclude that the transition dynamics between the states is consistent with a memoryless (continuous time) Markov chain whose transition probabilities are found in Fig. 6h and transition rates out of the closed, partial and open states are 4, 4.76 and 1.05 1/s, respectively.

Discussion

It is well known that moisture is a prerequisite for swarming in all bacterial species. This understanding, which has been noted since the first, pioneering works on bacterial swarming[4,60,61] concerns the overall environment of the colony. In other words, it is a global requirement.

In the current work, we address the spatial and temporal local conditions from cell movement within a swarm. We showed that while clusters of cells migrate rapidly on the agar, cells in the swarming state with no close neighbors do not move until they are covered with enough liquid. Our results show that the overall availability of water plays a significant role in determining the fraction of temporarily immobile cells among a population, and the duration each of these cells will be stationary. We also see that cells that were not trapped by the local depletion of liquids are not affected and will move with similar statistics, regardless of the overall availability of the water. In addition, we show that flagellar arrangement is strongly correlated with the cellular motion, which is also strongly correlated with the local presence of liquid around the cell. Cells that are temporarily left alone are immobile, are not covered with enough liquids, and have their flagella widely open. Cells in a cluster are mobile, are covered with liquid, and have their flagella arranged in a bundle. With that, our results cannot conclude on what is the cause and what is the effect. The transitions between the flagellar states are consistent with a continuous time Markov chain.

Previous works showed that during swarming, the chemotaxis mechanism is suppressed[19,62]. However, cells do change direction through spontaneous clockwise flagellar rotations. This is true even in cheB mutants (*B. subtilis* strain DS90) that are locked in the counter clockwise direction and are constantly in a run-only phase when in liquid bulk (but not on agar)[42,62]. Thus, the complete opening of the flagella while stalling is not necessarily related to chemotactic behavior. These dynamics are similar to previously reported observations by Turner et al[62], who noted that *Escherichia coli* swarm cells follow four kinds of tracks (modes) that strongly depend on the flagellar arrangement: stalls, reversals (not studied here), lateral movement, and forward movement. The "stall" mode, which is similar to the one reported here, was mostly found in cells that are at the edge of the colony. Moreover, these cells have their flagella completely spread out, pointing outward[63]. They speculated that this is a mechanism for extracting or pumping water from the substrate, thus moving the colony edge forward. Our results on *B. subtilis* show that the completely open flagella state is also observed inside the active swarm among solitary cells that have a temporal depletion of liquid around them. This suggests that the "stall" state may be related to a cellular mechanism for extracting liquid.

Overall, we showed that moisture and the local amount of liquid around cells play a pivotal role in understanding the biophysics underling the bacterial swarming phenomenon. Specifically, dry models of self-propelled agents, which only consider steric alignments and neglect hydrodynamic effect, are oversimplified and are not sufficient to describe swarming bacteria.

Acknowledgments

We thank Carmit Ziv for sharing the DIC microscope, Eric Clément and Yilin Wu for discussions. This project has received funding from the European Union's Horizon 2020 research and innovation program under the Marie Skłodowska-Curie grant agreement No 955910. Support for this work comes from National Institutes of Health grant R35GM131783 to DBK.

Author contributions

A.J, A.E, D.B.K, G.A, and A.B designed research, A.J, B.P.E, and A.B performed the experiments, S.O., A.E. and D.B.K provided the strains and mutants, A.J, B.P.E, A.E, D.B.K, G.A and A.B analyzed data, A.J, B.P.E, S.O.B, A.E, D.B.K, G.A and A.B wrote the paper.

Competing interests

The authors declare no conflict of interests.

Data availability:

All data will be available upon request.

Figures

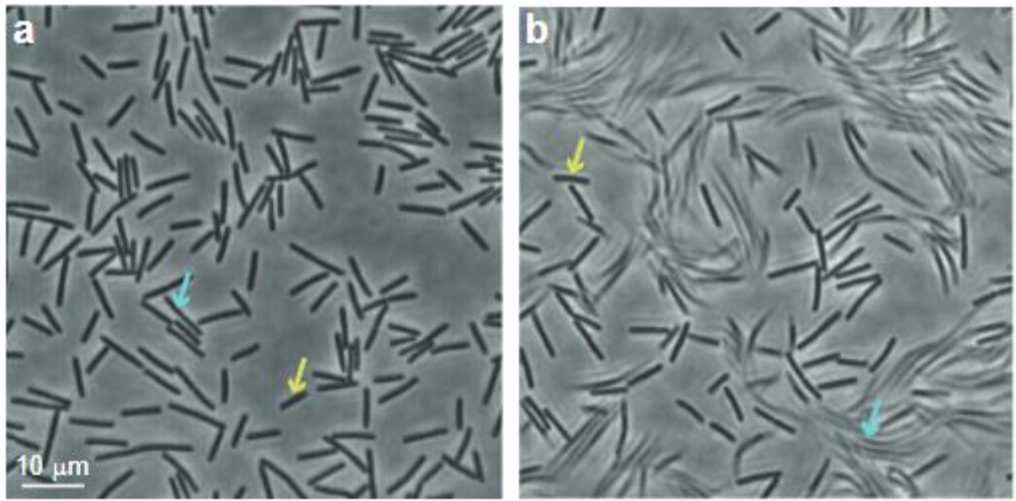

Fig. 1. **Phase-contrast microscopy of swarm cells on agar for two different exposure times.** (a) High frame rate (63×, 50 frames/s; exposure time=0.02 s). All the cells ($\rho$=0.3) are clearly observed. (b) At a slower frame rate (63×, 5 frames/s; exposure time=0.2 s), only cells that are temporarily immobile are sharp, while moving ones are smeared. Yellow arrows indicate temporarily immobile cells, and pale-blue arrows indicate moving cells.

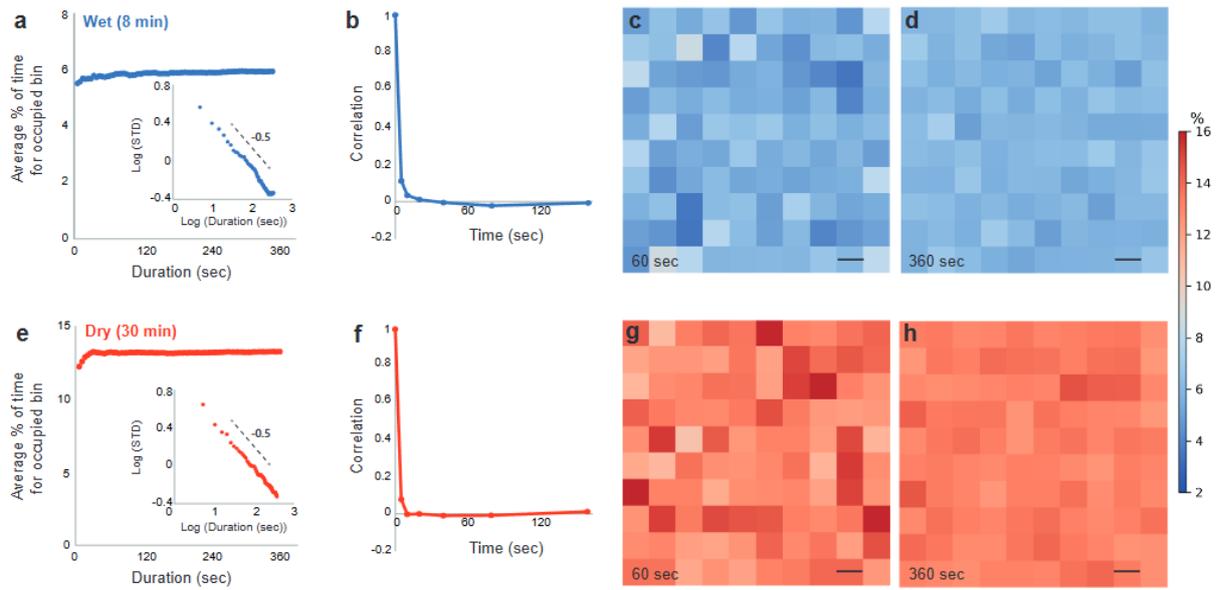

Fig. 2. **Inhabitation of "bins" (regions on the agar) by temporarily immobile cells is larger for dryer plates but over time the surface is statistically uniform.** (a) A wet case, following drying the plate for 8 min. The average % of time a bin (total number of bins in a frame = 100) is occupied is approximately 6%, regardless of the time interval (5-360 s). The inset shows that the standard deviation decreases as $t^{-1/2}$ with time. (b) The temporal correlation in the bin occupation decays to zero with a characteristic time of about 5 s. (c) An example of the different bins occupied by the temporarily immobile cells for the case of 60 s and (d) for 360 s. (e) A dry case, following drying the plate for 30 min. The average % of time a bin (total number of bins in a frame = 100) is occupied is approximately 13%, regardless of the time interval (5-360 s). (f) The temporal correlation in the bin occupation (dry case) decays to zero with a characteristic time of about 2 s. (g) An example of the different bins occupied by the temporarily immobile cells for the case of 60 s and (h) for 360 s. Scale bar in c, d, g, and h, is 10 μm.

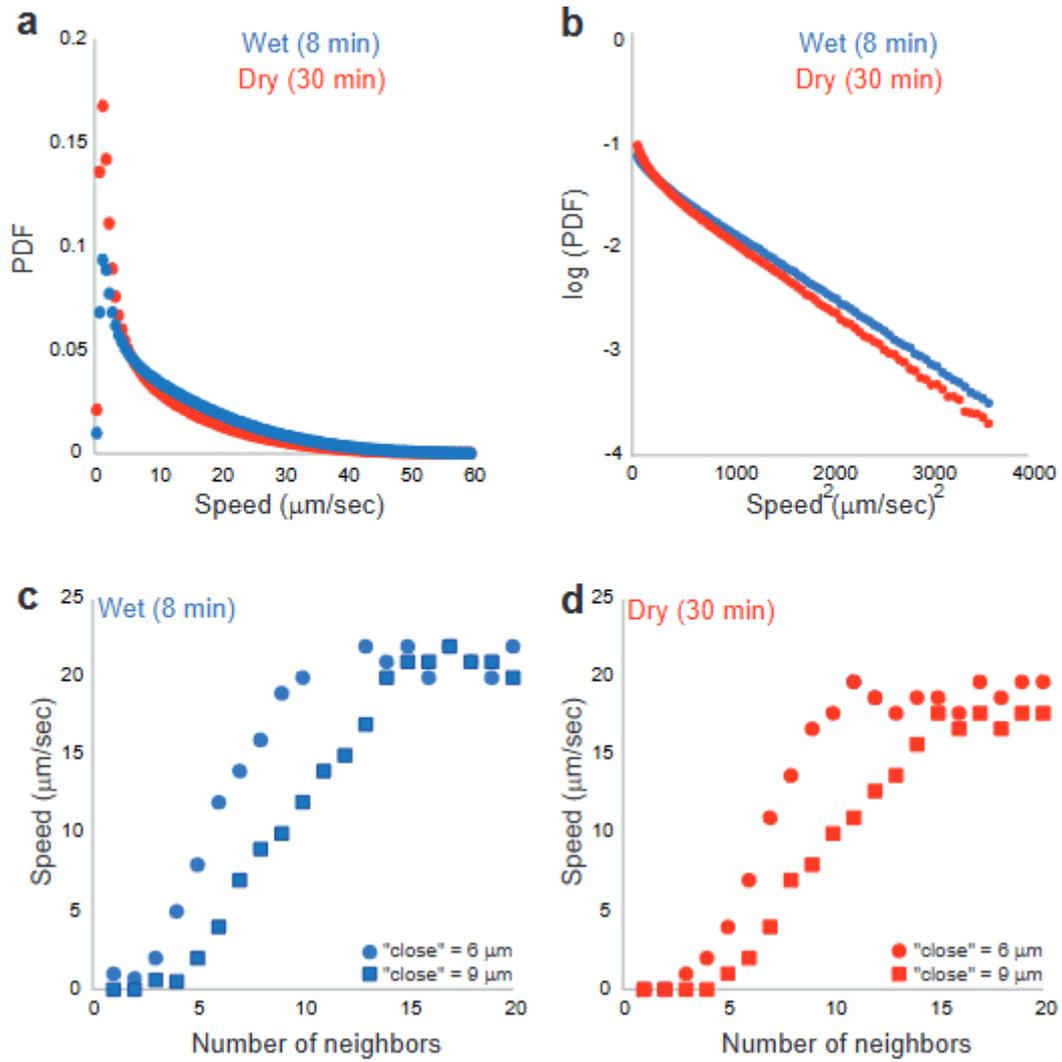

Fig. 3. **The distribution of speeds in the two (wet and dry) cases.** (a) Data presented for all speeds. Differences are mostly apparent at small speeds. (b) Data shown for speeds larger than 7 μm/s suggest similar distributions with Gaussian tails. (c-d) The average instantaneous speed of individual cells as a function of the number of their neighbors up to a cutoff distance "max distance", (c) for the wet and (d) dry cases. Cells that do not have any close neighboring cells do not move. Then, the average speed increases with the number of neighbors up to a maximum of approximately 20 μm/s.

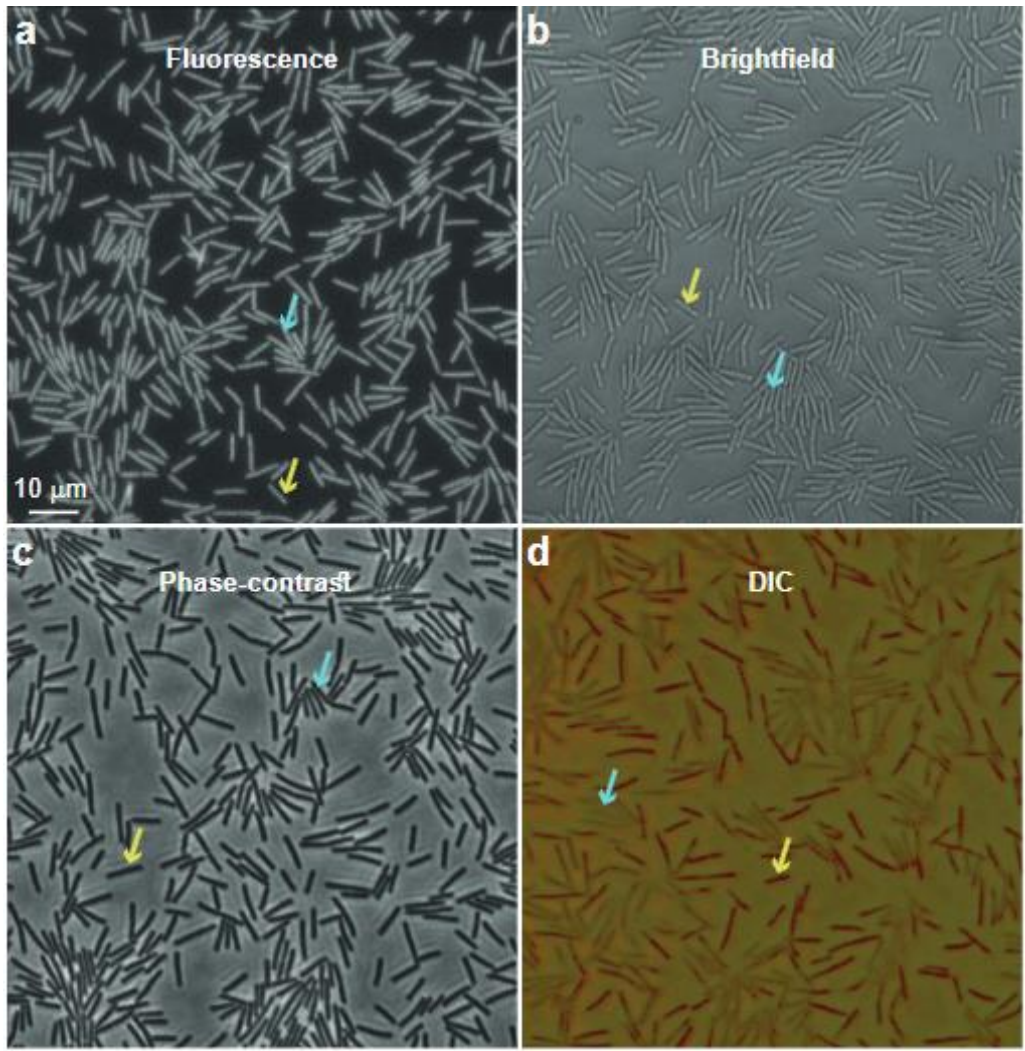

Fig. 4. **Indication for cell motion/immobility using high frame rate.** We compare four different techniques using (a) fluorescence, (b) brightfield, (c) phase-contrast, and (d) DIC. Yellow arrows indicate temporarily immobile cells, and pale-blue arrows indicate moving cells. Only DIC (d) reveals the temporarily immobile cells at even a high frame rate (50 frames/s; exposure time=0.02 s).

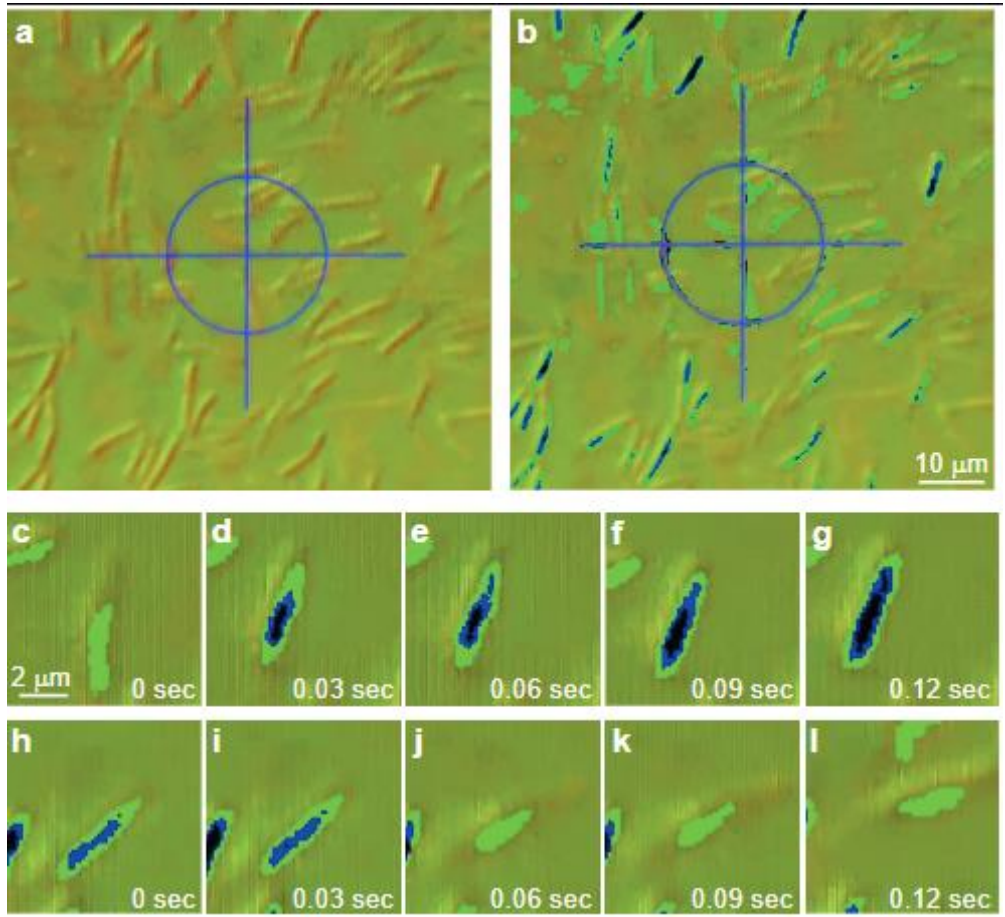

Fig. 5. **Transmitted-light DIC microscopy of swarming bacteria.** (a) The field of view shows different colors (hues) that depend on the topography of the specimen. (b) Each pixel in (a) gets a new easy-to-follow-color (green, blue, black) based on its original hue-value. (c-g) A moving cell stops, and its color changes from green to black, indicating that the original hue-value changes, suggesting that the topography of the sample is different due to the lack of liquid (time intervals from the first image are indicated in the other images). (h-l) A temporarily immobile cell starts moving, and its color changes from black to green, indicating that the original hue-value changes, suggesting that the topography of the sample is different due to the presence of liquid (time intervals from the first image are indicated in the other images).

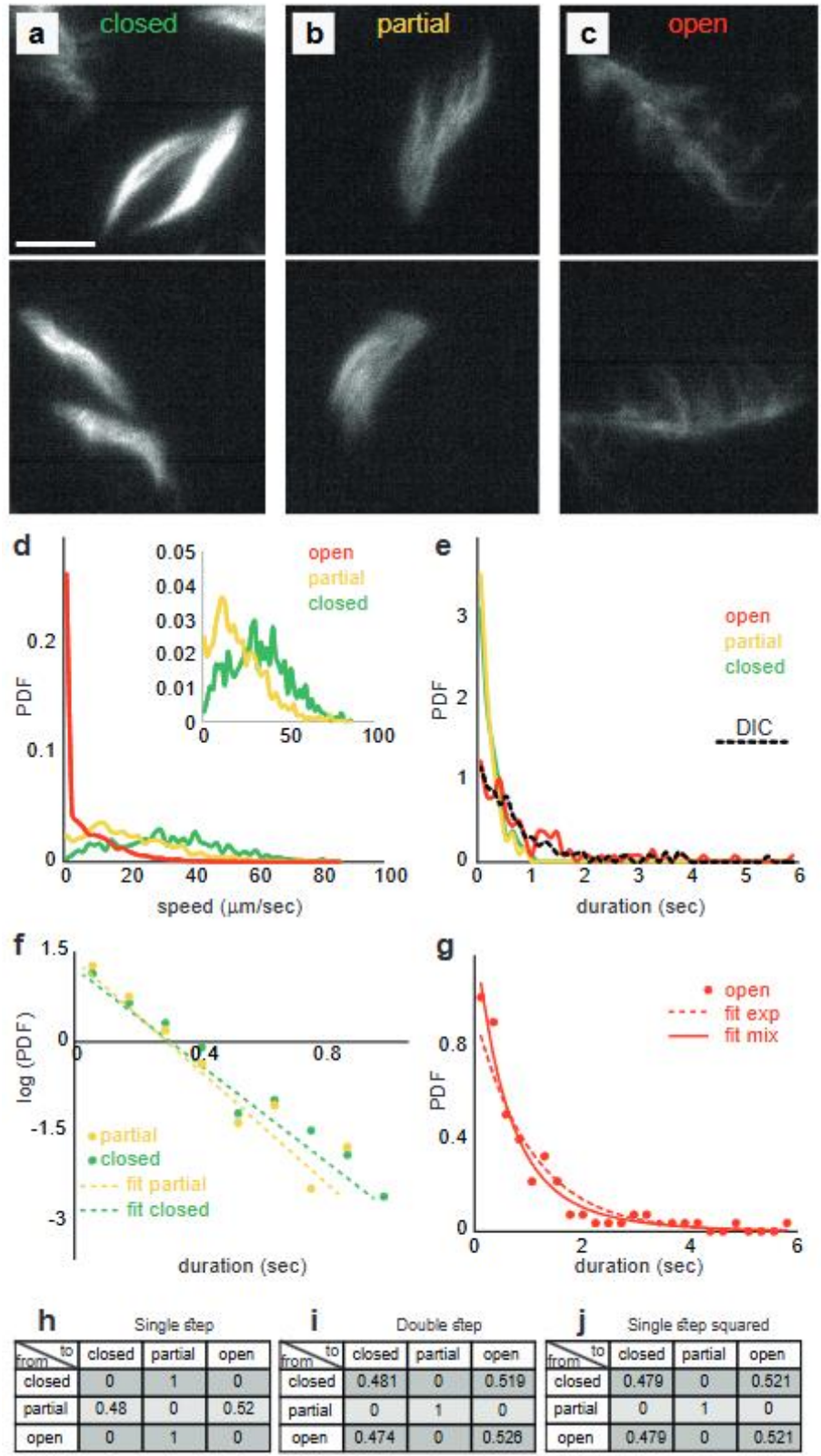

Fig. 6. **Analysis of the motion of swarm bacteria based on the stained flagella.** (a-c) Two examples; (a) Flagella are closed; all flagella tightly wrapped around one of the poles, (b) flagella are partially open (partially unbundled), and (c) flagella are open. Error bar equals 10

µm. (d) Correlation between flagellar arrangement and cell speed. The probability of speeds: cells with completely unbundled (open) flagella (data in red) are almost never moving, while cells in the two other states do move, but with different speed distributions (inset). (e) The distribution of time durations for which cells remain in a state. The distributions are practically the same for closed (green) and partially open (yellow) cells, but times are significantly longer for cells with open-flagella (red). Averages are 0.25 s for the closed, 0.21 s for the partial, and 0.95 s for the open. The distribution of the time duration in the open-flagella state, taken from the DIC experiment is superimposed (dashed black). (f) A histogram of the waiting times in the closed and partially-open states on a semi-log plot along with the maximum likelihood fit to an exponential distribution (averages 0.25 s and 0.21 s, respectively). (g) A histogram of the waiting times in the open state. The dashed and full lines show fits to a single and mixture of two exponential distributions (average 0.95 s). (h) The transition matrix between the states. (i) The two-step transition matrix. (j) The square of the single step one in (i). The two are equal up to a relative error of ~1%. Overall, we conclude that the transition dynamics between the states is consistent with a memoryless (continuous time) Markov chain.

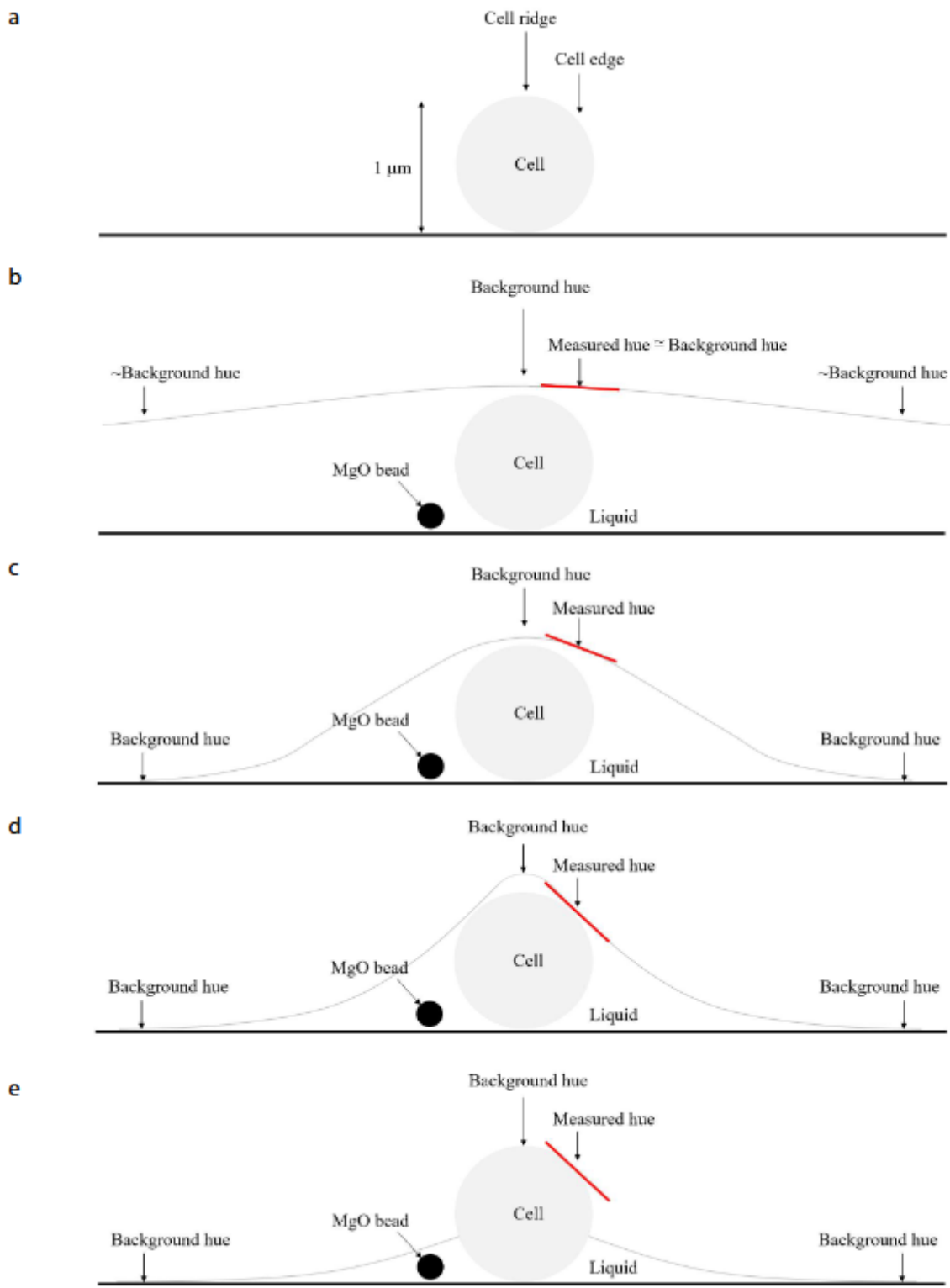

Fig. 7. **Experiments with silica beads.** (a) Phase-contrast microscopy of silica beads covered by a thin water layer. (b) DIC microscopy of silica beads covered by a thin water layer.

<u>Supplementary Information</u>

Movie S1. A real-time, phase-contrast movie of swarming cells near the colony edge, with tiny MgO smoke beads gently deposited on the colony. Beads outside the colony are immobile. Cell-size is ~1×7 μm.

Movie S2. A phase-contrast movie of swarming cells inside the colony (100 μm away from the edge), with tiny MgO smoke beads gently deposited on the colony. Beads move fast close to temporarily immobile cells, indicating that there is flowing liquid near the stationary cells. Cell size is ~1×7 μm; frame rate is reduced by a factor of 25.

Movie S3. A real-time movie showing fluorescence labelling of flagella in swarming cells. Only the flagella are labelled, the cell-bodies are invisible. The fraction of labelled cells is 1/20 (there are many more cells on the agar but they are not shown).

Appendix 1 – TL DIC microscopy

In the current work we use Transmitted-Light Differential Interference Contrast (TL DIC) microscopy equipped with color "enhancement". As in all color-DIC systems, the incident light travels through a polarizer, a Rochon/Wollaston/Nomarski prism that splits the ray, and a lambda plate. In our case, the light hits the sample from below, the bottom side of the sample is flat and smooth, and the substrate does not have birefringent properties (the bottom of the Petri-dish is made of a special glass and not the standard plastic).

Loosely speaking, the color-DIC system can differentiate between slopes in the sample, which are manifested in the hue property of colors. Differences in the index of refraction between the cells and the water (liquid), which are typically smaller than 0.05, do not play a role in determining the obtained hue, specifically in our systems where the cells are thin; Within the tested range, the same hue will be obtained for all objects in the sample if they have the same slope. In non-color DIC for instance, the differences in the index of refraction are resulted in an almost complete destructive interference, and only the perimeter of the cells may be resolved. This means that in color-DIC, cells embedded in water are almost invisible (in contrast e.g., to phase-contrast microscopy where cells embedded in water are nicely resolved).

Overall, within the experimental conditions and range of the slopes measured, the obtained hue is a monotonic function of only the local slopes in the upper surface and not of the material through which the light is transmitted. Mind that the saturation and luminance do depend on the material (its thickness or density for instance) through which the light is transmitted so the cells may sometimes be seen e.g., in Fig. 4c.

The lateral resolution (in the viewing plane) of DIC is of the order of the wavelength i.e., ~0.5 μm (stemming from the 0.2 μm adjutant points of the two split rays), and the ability to resolve the vertical axis by the hue is typically of the order of 0.05 μm, which yields an approximate minimal "local" angle of detection of ~4˚ (i.e., slopes smaller than 4˚ are not easily resolved; there is also a maximal angle above which the light is reflected away from the lens and the sample becomes dark).

In order to relate between the liquid around a cell and the measured hue value, consider a cross section of a bacterium with a cylindrical shape with a diameter 1 μm, resting on a flat surface (see Fig. 7a). Liquid (e.g., water) may cover the cell completely or partially. In Fig. 7b, the cell is completely covered. The height of the water is not important – only the angle it creates with respect to the tangent to the "ridge" of the cell. Thus, all samples that are covered by a flat-water interface will yield the same hue, regardless of the water height. In Fig. 7c we show a different case where the liquid does not completely cover the cell, resulting in different interfacial slopes around the cell. The ridge of the cell is parallel to the background and will yield the same hue. However, the edges of the cell are covered with liquid that creates a changing slope; the local slope will yield a hue that reflects the slope's value. In Figs. 7d-e, the layer of liquid is shallow, or there is no liquid at all. In these cases, only the ridge of the cell has a parallel surface that will yield the same hue as the one seen for the background. The edges of the cell will have different hues depending on their local slopes, independently of the liquid. Upon increasing the liquid layer, the system transitions from Figs. 7d-e to Fig. 7c or even Fig. 7b, and the slopes of the liquid on the cell edges become moderate, which changes the hue to the background value. Note that the ridge of the cell is very narrow; it is narrower than the lateral resolution – therefore, the background hue is seldom obtained on the ridge.

To illustrate how DIC works on transparent objects embedded in transparent medium, we performed an experiment described next and documented in Fig. 8. In Fig. 8a we show a set of images of spherical silica beads (5 μm in diameter) deposited on an agar surface. A drop of water, mixed with surfactant, is allowed to spread over the agar. The surfactant is added to form a relatively thin liquid front. The phase-contrast images are presented in order to show the advancing edge of the drop that covers the beads. The images show that the beads are nicely resolved with the water on top of them, and that they are not being flashed by the liquid (they are slightly embedded in the agar and a slight motion might be detected as the agar is rewetted). The focus is adjusted after the water covers the beads. The DIC set of images (Fig. 8b), shows a silica bead and the different colors (hues) that are exhibited along a diagonal line parallel to the beam splitting direction (from top-left to bottom-right). The hues reflect the structure of both the bead and the surface on which it is deposited. The top of the bead has the same hue as the one seen at the background (pale-blue). The intensity of the outer regions of the bead are low due to the relatively large angles, as most light is reflected away from the objective lens. The reddish and yellowish colors along the beam splitting direction, but outside of the bead, indicate that the bead is slightly embedded in the agar. When water approaches, the bead is completely covered with water, and the surface becomes flat and smooth and almost parallel to the agar (except for the advancing contact angle which is very small ~ 7˚; it thus appears brown and not pale-blue). In contrast to the phase-contrast case, in the DIC images the bead is completely invisible – even though we travel with the stage to find the appropriate focus. After a few minutes (the process is gradual but may be fairly seen after few minutes), the surface becomes pale-blue again because the angle of the drop decreases. The experiment lasts about 20 min during which the water evaporates and absorbed by the agar (and the bead drifts a bit with respect to the lab because the soft agar is not stable enough during the "flood"). During capturing, we play with the focus every few seconds to show that the bead cannot be detected. The bead sometimes appears as a pale ring with a dot at its center, but only when the water level is low enough, the colors reappear.

Appendix 2 - flagella staining

Our protocol is similar to other protocols recently used to stain flagellar filaments in *Bacillus subtilis* 3610 and other species[63,64]; yet, in our case the cells are swarmers. In the strain we have used (DS1916, amyE::Phag-hagT209C spec), the flagellin protein genes have been modified to include the cysteine amino acid that binds to the maleimide functional group, allowing staining using Alexa 546 dye (ThermoFischer ref A10258). Bacteria are grown overnight (~18 h) in 2 ml of LB (25 g/l), at 30°C and shaking (200 rpm), from which 10 µl is used to inoculate three swarm plates (LB with 0.5% agar). When the droplets dry out, the plates are placed in an incubator at 30°C with controlled high humidity (~95% RH). Additionally, a fourth plate is inoculated with a one-hour delay. This plate will be used to reincorporate the stained bacteria into swarming conditions. To facilitate that process, the inoculation is done slightly off-center. The remaining overnight culture is centrifuged for 3 minutes at 2000g, and the supernatant is saved for later.

The three initial plates are incubated for five hours, after which the bacteria are collected by flushing the colonial edge with phosphate buffer (0.01 M ph~7.2) and then collecting the liquid with the cells. Special attention must be put into collecting bacteria exclusively from the edge, where bacteria are in the swarming state. If not, bio-aggregates are created by bacteria in the biofilm, disturbing the following steps. After collection, flagella are stained by adding 5 µl of the dye solution (20 µg/µl of Alexa 546 in DMSO). We allow staining to occur for one hour while the suspension is rotated at 100 rpm to prevent bacterial sedimentation. Then, bacteria are centrifuged for 3 minutes at 2000g and resuspended in the overnight supernatant (0.2 µm filtered). Then, 20 µl is inoculated into the fourth agar plate. The inoculation drop is introduced outside of the colony, and then it spreads along the boundary of the colony, forming an arc, never being in direct contact with the colony. After drying, the plate is replaced in the incubator for 30 minutes or until the colony integrates with the stained bacteria. At the end of this process, one will obtain a swarming colony where a fraction of the cells has stained flagella.

Observation was done using the same fluorescence setup described above (Zeiss Axio Imager Z2 at 63×, filter set 20 Rhodamin shift free: Excitation 546/12; Beam Splitter 560; Emission 607/80), and a Zyla Andor operated at 100 frames per second.

Appendix 3 – image analysis

The quantitative analysis presented in Figs. 2 and 3 is facilitated using custom image analysis software implemented in Matlab. First, images undergo standard smoothing and preprocessing, as detailed in[41]. Applying a threshold on the intensity yields a Boolean mask indicating whether each pixel corresponds to a cell or not. Occupancy statistics (Fig. 2) corresponds to pixel-wise counts (and not number of cells).

The sequence of images is used to obtain velocity field using optical flow analysis (see[31] for details). Such algorithms inherently involve smoothing. Average speeds (Figs. 3ab) are obtained directly from the vector fields without taking into account speeds below a threshold. In other words, we average only over pixels that correspond to moving objects. The implementation follows[41].

Figures. 3cd show the average speed of cells as a function of the number of neighbors (up to a given distance). This requires identifying the position of individual cells and therefore tracking of individuals. To this end, we apply a novel method for tracking elongated objects using the Hough transform. First, the Boolean mask is separated into connected blobs using standard image analysis tools in Matlab. Blobs that are too small (in area or length) are discarded. The main difficulty is that cells that are practically touching are often associated with the same blob. To bypass this problem, in each blob, we find the longest line segment using the Hough transform. This segment (with some added thickness) is removed from the blob. Segments that are longer than 1.5 the average cell length are split in two. Repeating the process, we obtain a list of line segments that correspond to the individual cells.

Once individual cells were identified as line segments, we need to associate cells in consecutive frames to obtain trajectories. Typically, tracking software use the central points to find which cell in one image is closest to which cells in the following one. Here, we use the sum of the distances between the endpoints, which also takes into account the orientation and length of the detected cells. Since the order of the endpoints in each line segment is arbitrary, one needs to take the minimum between two ordering options. We perform a standard greedy pair matching algorithm, starting at the closed pair up to a cutoff distance of about 3 average cell lengths.

Next, the endpoints of each trajectory are smoothed over time using malowess with a linear interpolation. Velocities of the center of mass can then be calculated. Finally, to calculate the number of neighbors a focal cell has up to a given distance, we find the average number of endpoints associated with other cells to each of the endpoints of the focal cell. In other words, instead of counting cells, we count endpoints, each "worth" half a cell. This counting is used to calculate the statistics in Figs. 3cd.